\documentclass{PoS}

\usepackage{xspace}

\newcommand\ensuretext[1]{\ensuremath{\textnormal{#1}}}%
\newcommand\ergcms{\ensuretext{erg\,cm$^{-2}$\,s$^{-1}$}}%
\newcommand\ergs{\ensuretext{erg\,s$^{-1}$}}%
\newcommand\g{\ensuremath{\gamma}}%
\newcommand\fermi{\textit{Fermi}/LAT\xspace}%
\newcommand\fluxone{\ensuremath{(8.6 \pm 2.3) \times 10^{-12}}\,\ergcms}%
\newcommand\fluxtwo{\ensuremath{(1.6 \pm 0.3) \times 10^{-11}}\,\ergcms}%
\title{Seyfert~2 galaxies in the GeV band: jets and starburst}

\ShortTitle{Seyfert~2 galaxies in the GeV band: jets and starburst}

\author{\speaker{J.-P.~Lenain}, C.~Ricci, M.~T\"urler, D.~Dorner, R.~Walter\\%\thanks{A footnote may follow.}\\
        ISDC Data Centre for Astrophysics, Observatoire de Gen\`eve, Universit\'e de Gen\`eve, Chemin d'Ecogia 16, 1290 Versoix, Switzerland\\%
        E-mail: \email{jean-philippe.lenain@unige.ch}%
}

\abstract{The \fermi collaboration recently reported the detection of starburt galaxies in the high energy \g-ray domain, as well as radio-loud narrow-line Seyfert 1 objects, which were previously unusual extragalactic suspects for high energy emission. Motivated by the presence of sources close to the location of composite starburst/Seyfert 2 galaxies in the first year \fermi catalogue, we studied high energy \g-ray emission from such objects, aiming at disentangling the emission of starburst and Seyfert activity. We analysed 1.6 years of \fermi data from NGC\,1068 and NGC\,4945, which count among the brightest Seyfert 2 galaxies. We found an excess of high energy \g-rays of 8.3$\sigma$ and 9.2$\sigma$ for 1FGL J0242.7$+$0007 and 1FGL J1305.4$-$4928, which are found to be consistent with the position of the Seyfert 2 galaxies NGC\,1068 and NGC\, 4945, respectively. For both sources, we detect no significant variability nor any indication of a curvature of the spectrum. While the high energy flux of NGC\,4945 is consistent with starburst activity, that of NGC\,1068 is an order of magnitude above expectations, suggesting dominant emission from the active nucleus. We show that a leptonic scenario can account for the multi-wavelength spectral energy distribution of NGC\,1068. High energy \g-ray emission is thus revealed for the first time in a Seyfert 2 galaxy. If this result is confirmed in other objects, new perspectives would be opened up into the GeV band, with the discovery of a new class of high energy \g-ray emitters.
}

\FullConference{25th Texas Symposium on Relativistic Astrophysics - TEXAS 2010\\
		December 06-10, 2010\\
		Heidelberg, Germany}

\begin{document}

\section{Introduction}

The \fermi collaboration recently reported the discovery of four radio-loud narrow-line Seyfert~1 galaxies \cite{2009ApJ...699..976A,2009ApJ...707L.142A}. Previously undetected at high energies, these objects thus constitute a new class of high energy emitters. Extending this idea to Seyfert~2 galaxies, we searched for high energy emission from two active galactic nuclei (AGN) among the closest and brightest in the X-ray sky, NGC\,1068 and NGC\,4945.

NGC\,1068 is an archetypal Seyfert~2 galaxy, located at $z=0.003786$, i.e. 14.4\,Mpc away, harboring a hidden Seyfert~1 core. Antonucci \& Miller proposed the AGN unification based on studies of this source \cite{1985ApJ...297..621A}. Both AGN and starburst activities are present in the central region of NGC\,1068 \cite{1987ApJ...321..755L,2004Natur.429...47J}, the latter dominating the infrared emission of the broadband spectral energy distribution (SED) \cite{1989ApJ...343..158T}.

The presence of a \fermi source, 1FGL\,J0242.7$+$0007, in the region of NGC\,1068 in the 11-months \fermi catalogue (1FGL) \cite{2010ApJS..188..405A}, without any proposed counterpart in radio nor in \g-rays, motivated us to analyse 1.6 years of data from \fermi to better assess the origin of this \g-ray emission.

NGC\,4945, another Seyfert~2 galaxy at $z=0.001908$, also exhibits starburst activity within its core. Its emission was detected up to soft \g-rays, as observed with {\it INTEGRAL}/SPI, and was found to be Compton thick \cite{2009A+A...505..417B}. This source was reported as a \g-ray emitter in the 1FGL catalogue, although the authors did not conclude on the starburst or AGN origin of the emission.

\section{Data analysis}

We analysed $\sim$1.6\,yr of \fermi data, spanning from August 4, 2008 to March 15, 2010, from a region of interest of 10$^\circ$ in radius around NGC\,1068, using the publicly available Science Tools\footnote{\href{http://fermi.gsfc.nasa.gov/ssc/data/analysis/software/}{http://fermi.gsfc.nasa.gov/ssc/data/analysis/software/}}, and we followed the unbinned likelihood analysis scheme \cite{2009ApJ...697.1071A}.

Assuming a power-law shape for the source spectrum, the Test Statistic (TS) \cite{1996ApJ...461..396M} of the likelihood analysis is 68.6, corresponding approximately to a 8.3$\sigma$ source detection in the 100\,MeV--100\,GeV range. The best-fit location of the source is $\alpha_{J2000} = 2^\mathrm{h} 42^\mathrm{m} 46^\mathrm{s}$, $\delta_{J2000} = 0^\circ 2' 14''$ with an error circle radius of $6.1'$ (68\% confidence level, CL), and is fully compatible with the position reported in the 1FGL catalogue. Given the angular distance between the {\it Fermi} source 1FGL\,J0242.7$+$0007 and NGC\,1068, and the optical extension of $6.5'$ for the host galaxy, we propose that this \fermi source is actually associated with the Seyfert 2 galaxy NGC\,1068.

All the sources reported in the {\it Fermi}/LAT 11-months catalogue \cite{2010ApJS..188..405A} within a radius of 15$^\circ$ around NGC\,1068 were included in the likelihood analysis, and modelled with power-law spectra. For NGC\,1068, we obtain $F_\mathrm{100 MeV - 100 GeV} =$\fluxone\ and $\Gamma = 2.31 \pm 0.13$. We also tried to fit the data with a broken power-law or a log parabola, but this did not improve the likelihood. No significant variability was found in the data.

We followed the same procedure for the analysis of {\it Fermi}/LAT data in the region of NGC\,4945, for the source 1FGL J1305.4$-$4928. The likelihood analysis on NGC\,4945 results in a TS of 85.3 in the 100\,MeV--100\,GeV band, equivalent to a 9.2$\sigma$ detection. The nominal position of the \g-ray emission, located at $\alpha_{J2000} = 13^\mathrm{h} 05^\mathrm{m} 33^\mathrm{s}$, $\delta_{J2000} = -49^\circ 26' 44''$ with an error circle radius of $3.2'$ (68\% CL), is only $1.6'$ away from the position of NGC\,4945. Assuming a power-law shape on the source energy spectrum, a photon index of $\Gamma= 2.31 \pm 0.10$ and a flux of $F_\mathrm{100 MeV-100 GeV}=$\fluxtwo\ are found. As for NGC\,1068, the use of a broken power-law or a log parabola did not improve the likelihood, and no significant variability was found in the data, which are statistically consistent with a constant for both sources.

The {\it ISGRI} data were reduced using the {\it INTEGRAL} Offline Scientific Analysis software\footnote{\href{http://www.isdc.unige.ch/integral/}{http://www.isdc.unige.ch/integral/}} version 9.0. We have also checked the \textit{Swift}/BAT \cite{2005SSRv..120..143B} spectra of the two sources. The 18 months \textit{Swift}/BAT spectra have been extracted from the BAT archive \cite{2010A+A...510A..47S} served by the HEAVENS source results archive (Walter et~al., in prep.).

\section{Discussion}

\begin{figure}
  \centering
  \includegraphics[width=0.8\columnwidth]{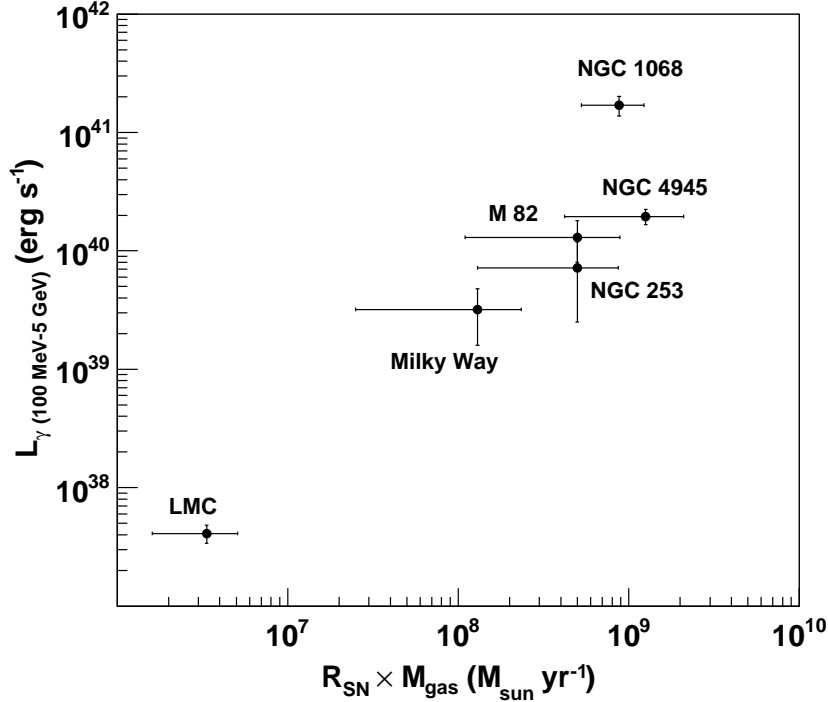}
  \caption{Relationship between SN rate, total gas mass and \g-ray luminosity of NGC\,1068, NGC\,4945, NGC\,253, M\,82, the LMC and the Milky Way.}
  \label{fig-SNrateMgas_GammaLumin}
\end{figure}

\begin{figure}
  \centering
  \includegraphics[angle=-90,width=0.8\columnwidth]{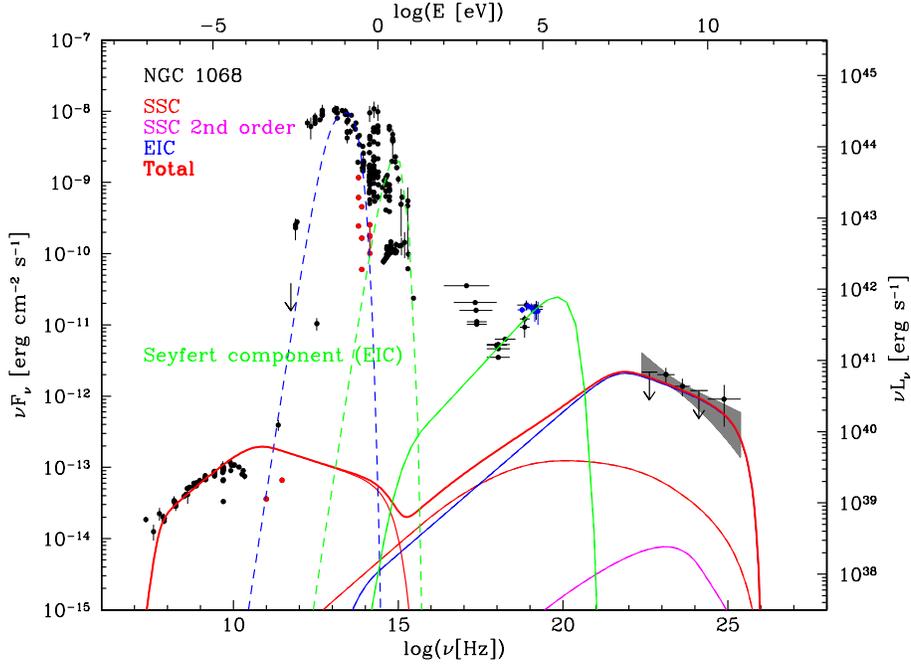}
  \caption{Spectral energy distribution of NGC\,1068, including the \fermi spectrum. The black and red points are archival data from the NED, the red ones denote data taken from the central region of NGC\,1068. For clarity, we only show the \textit{INTEGRAL} IBIS/ISGRI data in blue in the hard X-rays. The EIC model for the outflow is shown in blue, and the corresponding SSC emission is shown in thin red and magenta lines for first and second order components, respectively. The thick red line shows the sum of the different emission components from the large outflow. The EIC component from the accretion disc is shown in green.
  }
  \label{fig-model-NGC1068}
\end{figure}

As a first test to know whether the high energy emission of these two objects orginates from the starburst or the AGN component, a significant detection of variability would rule out the starburst hypothesis. However, due to the lack of statistics, no conclusions can be drawn about the variability of NGC\,1068 or NGC\,4945 from their light curves.

We also compare their \g-ray luminosity with those of the famous starburst galaxies NGC\,253 and M\,82, both detected at high energies with \fermi \cite{2010ApJ...709L.152A} and at very high energies with H.E.S.S. \cite{2009Sci...326.1080A} and VERITAS \cite{2009Natur.462..770V}, respectively. Both have a luminosity in the 100\,MeV--5\,GeV band of the order of $\approx$10$^{40}$\,\ergs, as detected with {\it Fermi} \cite{2010ApJ...709L.152A}. Computing the luminosities of NGC\,1068 and NGC\,4945 in the same energy band, for comparison, we obtain $1.7 \times 10^{41}$\,\ergs\ and $2.0 \times 10^{40}$\,\ergs, respectively. High energy emission from starburst galaxies is thought to originate from the interactions of cosmic rays produced by supernov\ae\ with the ambient interstellar medium, and the \g-ray luminosity $L_\g$ should then scale with the product of the supernova rate $R_\mathrm{SN}$ and the total gas mass $M_\mathrm{gas}$. We thus compare $R_\mathrm{SN}$, $M_\mathrm{gas}$ and $L_\g$ of NGC\,1068 and NGC\,4945 to the ones of NGC\,253, M\,82, the Large Magellanic Cloud (LMC) and the Milky Way (see Figure~\ref{fig-SNrateMgas_GammaLumin}), as well as their infrared and radio luminosities (see \cite{2010A&A...524A..72L} for more details). These objects are the only extragalactic sources which are not AGN known to emit high energy \g-rays. The \g-ray luminosity and the SN rate of NGC\,4945 are fully consistent with those of NGC\,253 and M\,82, hence even though this object is a composite starburst/AGN, its high energy \g-ray emission could be explained only in terms of starburst activity.

However, in the case of NGC\,1068, $R_\mathrm{SN}$ is comparable to those of M\,82 and NGC\,253, but its radio and \g-ray luminosities are higher by a factor $\sim$10. This would suggest that its high energy \g-ray emission is more likely dominated by the central AGN activity. This is also strengthened by the fact that radio maps of NGC\,1068 clearly show a structured jet, on parsec- and kiloparsec-scales, shaped by the outflow from the central AGN \cite{2004ApJ...613..794G,2006AJ....132..546G}. On the contrary, the radio morphology of NGC\,4945 shows an extended emission consistent with the optical morphology tracing the edge-on galaxy, consistent with a starburst emission.

Assuming that the \g-ray emission of NGC\,1068 is due to the central AGN, we apply an emission model developed for AGN which jet is misaligned to the line of sight \cite{2008A+A...478..111L}. A large, mildly-relativistic zone of the wind-like outflow, at a few tenth of parsecs from the core, could emit high-energy \g-rays through external inverse Compton process (EIC) \cite{1987ApJ...322..650B}. At such distances, the infrared photon energy density is still high enough to ensure a significant EIC emission while being not too important to prevent high optical opacity from pair production. In accordance with the data, no significant short-term variability is expected from such a large emitting zone.

In the model, a compact, dense plasma radiates through synchrotron self-Compton (SSC) and EIC processes. Second order SSC emission is also accounted for. Figure~\ref{fig-model-NGC1068} shows the results of our model, with synchrotron emission responsible for the radio emission, while the \fermi data are interpreted as EIC emission with the infrared emission providing the seed photons, from a multi-temperature blackbody. We also show the contribution from the SSC process, which is negligible compared to the EIC emission at the highest energies (see \cite{2010A&A...524A..72L} for more details). The hard X-ray spectrum is due to EIC radiation from another relativistic population of leptons, within hot plasma located in the vicinity of the accretion disc. It should be noted that imaging atmospheric \v{C}erenkov telescopes have the ability to strongly constrain the highest energy part of the particle energy distribution, and hence provide insights on the acceleration processes at work in this object. If the leptonic population of particles extend at higher energies, a significant signal should be detectable from NGC\,1068 with H.E.S.S., VERITAS, MAGIC or the future CTA (\v{C}erenkov Telescope Array) observatory \cite{2010arXiv1008.3703C}.

\begin{figure}
  \centering
  \includegraphics[angle=-90,width=0.8\columnwidth]{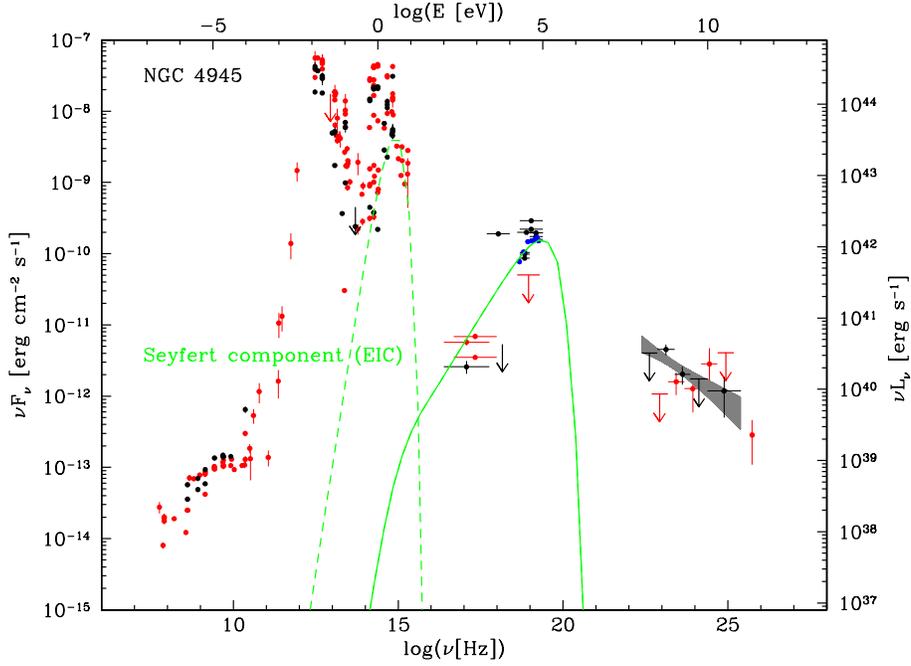}
  \caption{Spectral energy distribution of NGC\,4945, including the \fermi spectrum (black points). For clarity, we only show the \textit{INTEGRAL} IBIS/ISGRI data in blue in the hard X-rays. The model for the EIC component from the accretion disc is shown in green. We show in red the data of NGC\,253 as taken from the NED, with its \fermi spectrum \cite{2010ApJ...709L.152A} as well as the H.E.S.S. flux measurement \cite{2009Sci...326.1080A}, for comparison. The luminosity axis on the right is given for NGC\,4945.
  }
  \label{fig-NGC4945_SED}
\end{figure}

Figure~\ref{fig-NGC4945_SED} shows the SED of NGC\,4945 including archival data, as well as the \fermi and {\it INTEGRAL} spectra. We also show for comparison the SED of the starburst galaxy NGC\,253. The two sources clearly have very similar broadband SEDs, strengthening the idea that the high energy \g-ray emission from NGC\,4945 is of starburst origin. NGC\,253 being detected at very high energies with H.E.S.S. \cite{2009Sci...326.1080A}, NGC\,4945 is thus expected to be also detectable in this energy range.

\section{Conclusion}

High energy \g-ray emission is revealed and associated to the starburst/Seyfert~2 objects NGC 1068 and NGC\,4945, using \fermi data. Both source spectra are consistent with a power-law, and no variability is found in the data. Compared to M\,82 and NGC\,253, which high-energy emission is dominated by starburst activity, we find a too high \g-ray luminosity of NGC\,1068 to be explained only by starburst activity. We thus propose a leptonic scenario to interpret the high energy emission in terms of external inverse Compton process from an outflowing relativistic wind launched by the central AGN.

If high energy \g-ray emission due to AGN activity is confirmed in other Seyfert 2 galaxies, this would mark the discovery of yet a new class of high-energy \g-ray emitters. New data from \textit{Fermi}/LAT in the coming years will be extremely helpful to search for such emission.

\acknowledgments

This research has made use of NASA's Astrophysics Data System (ADS), of the SIMBAD database, operated at CDS, Strasbourg, France, and of the NASA/IPAC Extragalactic Database (NED) which is operated by the Jet Propulsion Laboratory, California Institute of Technology, under contract with the National Aeronautics and Space Administration.

\end{document}